# The yield normal stress

**Henri de Cagny[1], Mina Fazilati[1], Mehdi Habibi[1,2], Morton M. Denn[3], Daniel Bonn[1]**

[1] Institute of Physics, University of Amsterdam, Science Park 904, 1098XH Amsterdam, The Netherlands

[2] Laboratory of Physics and Physical Chemistry of Foods, Wageningen University, 6708WG Wageningen, The Netherlands

[3] Benjamin Levich Institute and Department of Chemical Engineering, The City College of New York, CUNY, New York, NY 10031, USA

## Abstract

Normal stresses in complex fluids lead to new flow phenomena because they can be comparable to or even larger than the shear stress itself. In addition, they are of paramount importance for formulating and testing constitutive equations for predicting non-viscometric flow behavior. Very little attention has so far been paid to the normal stresses of yield stress fluids, which are difficult to measure. We report the first systematic study of the first and second normal stress differences, $N_1$ (>0) and $N_2$ (<0), in both continuous and oscillatory shear of three model yield stress fluids. We show that both normal stress differences are quadratic functions of the shear stress both above and below the shear yield stress, leading to the existence of a yield normal stress.

Normal stresses are ubiquitous in complex fluids. They have been studied in great detail for polymer melts and solutions, where they emerge because of the stretching of the polymer chains [1]. The interest in their study has two main reasons. First, these elastic stresses are responsible for a number of spectacular flow effects such as the tubeless syphon, rod climbing and die swell [1]. Second, they have been instrumental in developing constitutive equations that provide general relations between the applied stress and the resulting deformation rate for general flows [1]. As a direct consequence, the prediction of the flow behavior of polymeric fluids in any type of flow has become possible, which has greatly advanced the field and is invaluable for the many applications of polymers.

To the contrary, no generally accepted constitutive relations are available for the very important class of yield stress fluids. Yield-stress fluids are ubiquitous in processes ranging from the extraction of oil to the production of personal care products and food [2] [3]. They are defined as materials that undergo a transition from a solid-like to a liquid-like state at a critical stress or strain. Yield stress fluids have been studied for roughly a hundred years, since the early work of Bingham. However, for most of this time the focus has been on measuring the shear stress as a function of shear rate, establishing the magnitude (or even existence [4]) of the yield stress, and on ways to quantify the shear rheology. From such measurements, yield stress fluids are typically described phenomenologically as Herschel-Bulkley fluids for which the shear stress $\tau$ depends on the shear rate $\dot{\gamma}$ as $\tau = \tau_y + K\dot{\gamma}^\beta$, where $\tau_y$ denotes the yield stress. Yet, this is far from a complete description of the material behavior [5]: it is now well accepted that common yield-stress fluids are viscoelastic both before and after yielding: they should be described as elasto-viscoplastic materials, including also a viscoelastic description of the normal stresses.

There has been considerable activity in recent years in trying to establishing invariant elasto-viscoplastic constitutive equations for yield-stress materials that predict the stress tensor for arbitrary deformations imposed on the material. These are typically formulated as generalizations of equations that have been developed successfully for polymeric liquids [3] [6] [2] [7] [8] [9] [10]. However, they are difficult to test without meaningful measurements of components of the stress tensor other than the shear stress; normal stress measurements are in fact standard in polymer rheology to test constitutive relations. Not only do very few measurements of the normal stress in yield-stress fluids exist, but those that exist do not provide a coherent picture: positive $N_1$ [11] [12] [13] [14] has been reported for some systems, whereas negative $N_1$ has been reported for others [15] [16]. The main issue that has to be dealt with is that their measurement is challenging because the flow is often heterogeneous [17] [18], and the systems may have residual stresses and uncontrolled trapped strains [19], underlining the importance of experimental protocols [14]. Furthermore, commercial rheometers provide in general only the average value of the normal stress during the duration of a transient or cyclic measurement, making it difficult to disentangle the shearing contribution to the signal from other phenomena, such as residual normal forces due to trapped stresses, white noise, and baseline drift.

In this article we report measurements of the two normal stress differences in three typical yield stress materials under continuous shearing and in slow oscillatory flow. We circumvent the problems due to the averaging and drifting of force measurements by recording the normal force signal directly: we connect an oscilloscope to the electric output of the rheometer and record the full history of the signal. It is then much easier to analyze the contribution of the shear to the normal force signal, for instance by applying a slow oscillatory stress and measuring the amplitude of the oscillatory output. We show that such slow oscillatory measurements provide a reliable determination of the normal stresses without introducing edge failure at large stress in the usual continuous shear flow. For three typical yield stress materials, we find that $N_1$ is positive and $N_2$ is negative and smaller but of comparable magnitude. Furthermore, both $N_1$ and $N_2$ follow a quadratic evolution with the shear stress that is continuous both above and below the yield stress, leading to the emergence of a yield normal stress.

We use a castor oil-in-water emulsion [20] and two polymer microgel suspensions: Carbopol in water and a commercial hair gel. These materials are known to be simple (non-thixotropic) yield stress fluids [21] and are widely used as model fluids [20] [22]. The emulsion is composed of 80% oil in water stabilized by sodium dodecyl sulfate (SDS), with a mean droplet diameter of about 5 micrometers. This emulsion is stable for months. The Carbopol gel is prepared by mixing 2 wt.% Carbopol Ultrez U10 and distilled water for one hour, after which 18 wt.% sodium hydroxide solution is used to adjust the pH to approximately 7. Finally, the Carbopol-water mixture is diluted to 0.7 wt.% Carbopol by adding distilled water. Hair gel is a commercial product available from supermarkets and consists of Carbomer stabilized by triethanolamine. From a microscopic point of view, theory suggests [23] [24] that similarly to the polymers, drop deformation in flowing emulsions leads to finite normal stress differences; the microgel particles of the other two systems are also soft and deformable, and should consequently behave similarly.

Rheological measurements are carried out with an Anton Paar MRC 302 rheometer with rough surfaces to avoid wall slip using a 50-mm diameter cone-plate (CP-50) geometry (1° cone angle) and a 60-mm diameter plate-plate (PP-60) geometry at a gap spacing of 1 mm. To have more sensitivity of the normal force $F$, additional experiments are performed using a home-made 125-mm cone-plate geometry (CP-125, 4° cone angle); for the CP measurements $N_1 = \frac{2F}{\pi R^2}$, and for the PP geometry the shear and normal stresses corresponding to the shear rate at the rim $\dot{\gamma}_R$ are calculated using:

$$\tau = \frac{M}{2\pi R^2}\left(3 + \frac{d \ln M}{d \ln \dot{\gamma}_R}\right)$$

$$(N_1 - N_2)_{\dot{\gamma}_R} = \frac{F}{\pi R^2}\left(2 + \frac{d \ln F}{d \ln \dot{\gamma}_R}\right)$$

Here, $M$ is the torque and $R$ is the radius of the plate. Combining CP and PP measurements therefore allows in principle to obtain both $N_1$ and $N_2$.

We first measure the time-resolved shear and normal stress response in stress-controlled oscillatory shear using an oscilloscope coupled to the analogue outputs of the rheometer. A prior calibration is made to convert the electric signal (originally in Volts) to relevant units. A set of weights is used to calibrate the normal force signal, and measurements on standard silicon oils are performed to calibrate the shear stress and strain signal. All mesurements are at a frequency of 0.1 Hz, which is sufficiently slow to permit stresses to relax and map out the steady-state response. Fig. 1(a) shows an example of the raw data for the oscillation measurement at an amplitude of 100 Pa, which is a strain outside the linear regime and past yielding. It is important to note that all measurements are recorded after initial transients due to stored and trapped stresses have relaxed and the shear and normal stresses from successive deformation cycles reach a steady state; this can take as long as 15 minutes for the smallest deformations. There is a small amount of drift in the normal stress transducer, and normal stress measurements are shifted uniformly to ensure that the minimum normal stress was equal to zero.

We construct flow curves (stress vs. shear rate) from the peak values of the oscillatory sweeps at constant frequency $\omega$. If the output strain amplitude is sinusoidal then the shear rate is calculated as $\dot{\gamma} = \gamma\omega$. As can be seen in Fig. 1a, the output strain deviates slightly from a pure sinusoid at high stress amplitude. We therefore performed a Fast Fourier Transform on the signals at high imposed shear stress, and when the contribution of the third harmonic to the shear rate became higher than 5%, we modified the value of the shear rate accordingly, as described in the Supplementary Information. This calculation was required for only a few points. This allows to directly compare the amplitude of oscillatory measurements to the usual Herschel-Bulkley flow curve obtained in continuous shear. As shown in Fig. 1, the agreement between the two types of measurements is excellent, and all data are well described by the Herschel-Bulkley model (for fit parameters see SI).

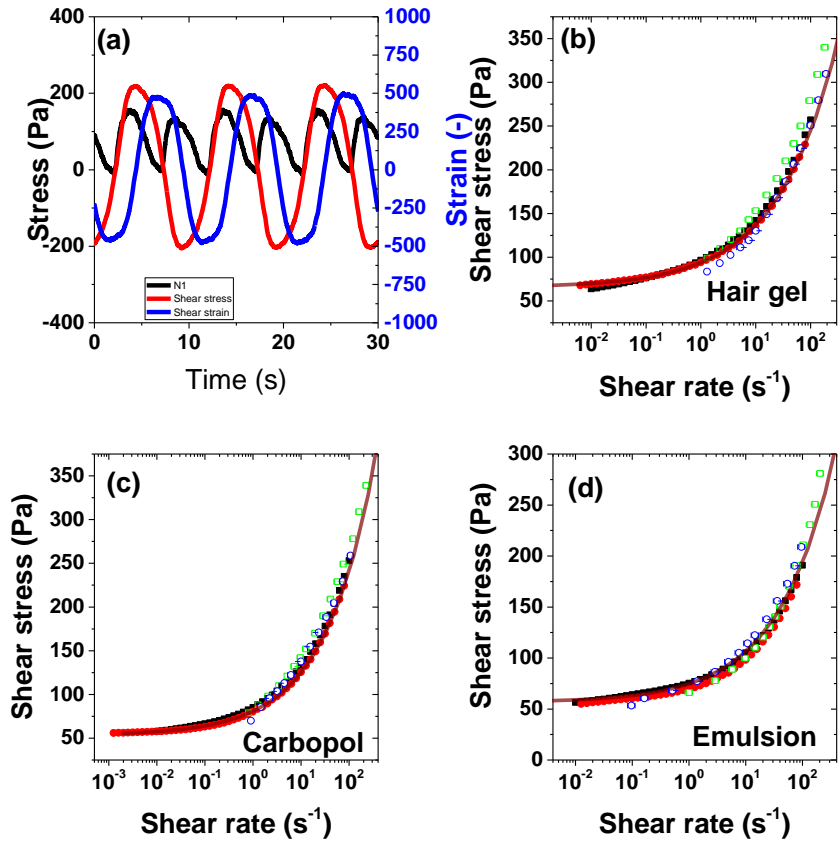

Figure 1: (a) : Time series of shear stress, shear strain, and normal stress recorded when applying a sinusoidal shear stress (frequency 0.1Hz, amplitude 210Pa) on a hair gel sample. All the electrical signals are smoothed using the Savitsky-Golay method *[25]* before measuring the amplitude of the oscillations. The uncertainty due to the white noise is estimated to be about 3 Pa. (b)(c)(d) Comparison of the flow curves from oscillatory and continuous shear, and from cone-plate ('CP') and plate-plate ('PP') measurements in oscillation for different yield stress fluids. The filled symbols represent rotation measurements while the open symbols are associated with the amplitude of the oscillatory measurements. The square symbols correspond to Cone-plate measurements and the circles to Plate-plate measurements. The line represents the Herschel-Bulkley fit. The samples are subjected to a pre-shear at a shear rate of 100 s⁻¹ for 30 s prior to each measurement series in order to remove any residual normal forces due to loading, after which we waited 30 s for normal forces to relax.

Having established the equivalence between the results of steady and oscillatory shear, Fig. 2 shows that $N_1$ and $N_1$–$N_2$ can be obtained reliably and reproducibly through imposing a slow oscillatory stress , recording the resulting electric signal on the oscilloscope, and measuring the amplitude of the normal stress oscillations. The log-log plot of Fig. 2 reveals that $N_1$ and $N_1$-$N_2$ for all three materials follow a quadratic relation $N_{1,2} = \psi_{1,2}\tau^2$; the corresponding normal stress coefficients $\psi_1$ and $\psi_2$ are given in Table 1. In the SI we show that the quadratic relation also applies within a single cycle, confirming that the slow oscillatory shear maps out the steady flow curve. It follows that for all systems $N_1$ is positive and $N_2$ is negative and somewhat smaller in magnitude than $N_1$, agreeing with simulation results for similar soft deformable particles by Seth *et al.* [24]. In Fig. 2, we also include the results obtained from classical rotational measurements from high to low shear rate, as recorded by the rheometer software. The oscillatory method appears to give better results; only at high rates do the experimental curves superimpose, and as the shear rate decreases, the noise in the continuous flow measurements and the discrepancy with the oscillatory measurements increase.

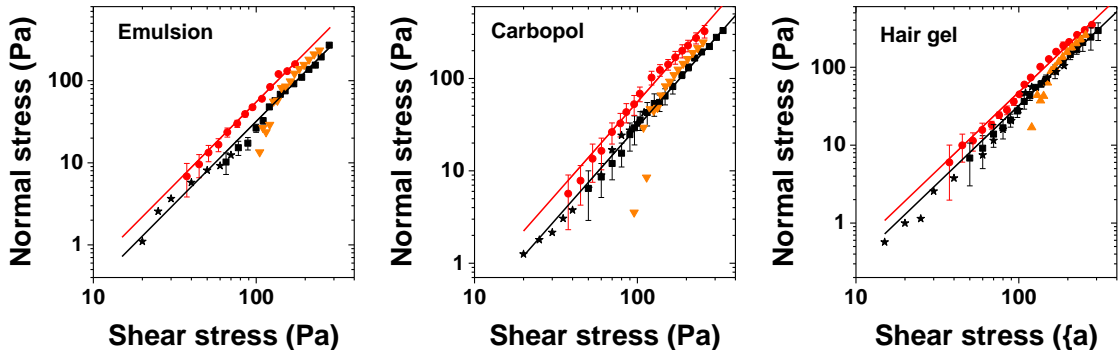

Figure 2: Normal stress measurements $N_1$ obtained with cone plate (black squares: CP-50; black stars: CP-125) and $N_1 - N_2$, obtained with plate-plate geometries (red circles: PP-60) for the three yield stress fluids. Each measurement was performed twice and the results averaged. The lines represent quadratic fits; the values of the slope of the fits $\Psi_1$ and $\Psi_2$ are listed in Table 2. The orange triangles represent $N_1$ from the steady shear experiment as reported directly by the rheometer software during a stress ramp.

The oscillation measurements discussed above are all done beyond the yielding point, because the amplitude of the normal stress at low shear stress is too low to be measured for stresses close to yielding. However the behavior in the vicinity of the yield point is of paramount importance: the yielding or not of these materials is at the origin of many processing, mixing and flow heterogeneity challenges that these materials pose. We therefore investigate the normal stresses close to yielding using the very large 125 mm diameter cone-plate geometry to have more sensitivity on the normal stress signal. Repeating the oscillation measurements we can now detect the normal forces for shear stresses as low as 20 Pa, well below the yield stress, which is in the range 50-80 Pa for the three systems studied here. Fig. 2 shows that the measurements done with the CP-50 and CP-125 cones agree well, and that even below the yield point the normal stress has the same quadratic dependence on the shear stress.

| Sample | $\psi_1$ (Pa$^{-1}$) | $\psi_1 - \psi_2$ (Pa$^{-1}$) | $\psi_2$ (Pa$^{-1}$) | Contribution to von Mises criterion | 2/G' (Pa$^{-1}$) |
|---|---|---|---|---|---|
| Emulsion | $3.19\cdot10^{-3}$ | $5.51\cdot10^{-3}$ | $-2.32\cdot10^{-3}$ | 0.4% | $2.2.10^{-3}$ |
| Hair gel | $3.22\cdot10^{-3}$ | $4.79\cdot10^{-3}$ | $-1.57\cdot10^{-3}$ | 0.4% | $5.0.10^{-3}$ |
| Carbopol | $2.99\cdot10^{-3}$ | $5.57\cdot10^{-3}$ | $-2.58\cdot10^{-3}$ | 0.6% | $5.3.10^{-3}$ |

Table 1: Coefficients obtained from quadratic fits to the normal stress versus shear stress curves $N_{1,2} = \psi_{1,2}\tau^2$ plotted in Fig. 2. The coefficient 2/G' has been added for comparison with $\psi$

Since the normal stress grows quadratically with the shear stress, the normal stress must tend to a non-zero value equal to $\psi_{1,2}\tau_y{}^2$ as the shear rate goes to zero: our results imply that the normal stress goes to a finite yield value for zero shear rate. The only way of establishing the behavior of the normal stress in the vicinity of yielding is to be able to impose very low shear rates using the CP-125. As a shear rate sweep leads to uncontrolled trapped stresses, we perform experiments in which we first impose a low shear rate value (ranging from $10^{-4}$ s$^{-1}$ to 1 s$^{-1}$), wait for a steady state, and subsequently set the shear stress to zero. Fig. 3a shows the results for one such test: the normal stress jumps from its value at the imposed shear rate to zero when the stress and shear rate simultaneously go to zero. These measurements allow one to measure $N_1$ for the three yield stress fluids over a large shear rate range; Fig. 3b also shows the results from the previous oscillation experiments with the CP 60 for comparison. These data show unambiguously that in addition to a shear yield stress there also exists a normal yield stress. The continuous red line follows from combining the Herschel-Bulkley stress $\tau = \tau_y + K\dot{\gamma}^\beta$ with the $N_1 = \psi_1\tau^2$ relation obtained in figure 2. There is good agreement using the same parameters, allowing only the prefactor k to vary slightly; this shows showing once again that the normal stress approaches a finite normal yield stress: $N_{1,y} = \psi_1 \tau_y{}^2$ .

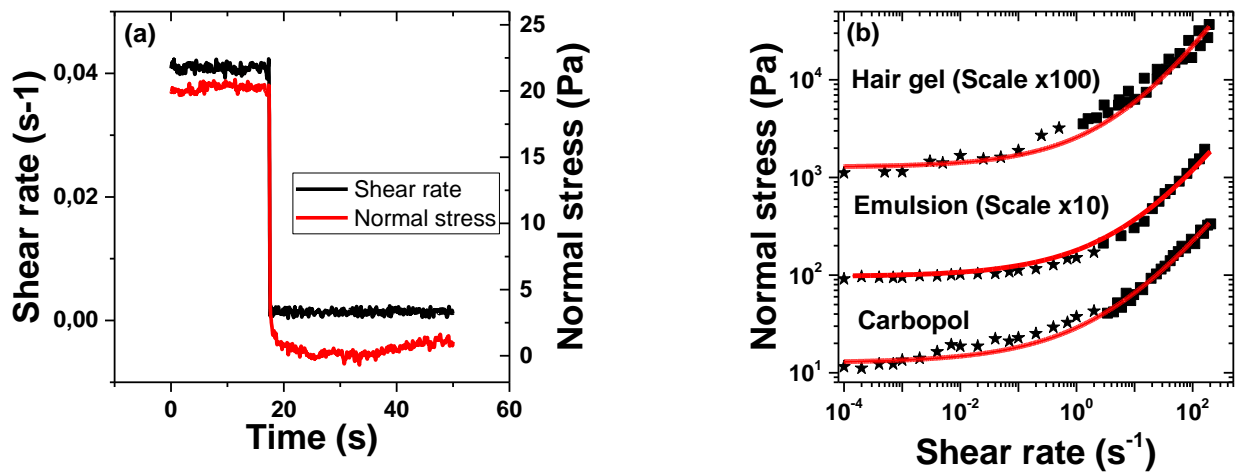

Figure 3: (a) Time response of the normal stress jump when imposing a zero shear stress after a steady state constant shear rate (here 0.04s$^{-1}$) . (b) $N_1$ as a function of shear rate for the three materials. Data (stars: CP-125; squares: CP-50) are shifted in the log-log plot for clarity. The red lines represent the fits obtained by combining the Herschel-Bulkley parameters from the SI with the quadratic normal stress/shear stress fits reported in table 2.

The quadratic dependence of $N_1$ on the shear stress is a characteristic of simple viscoelastic fluid models, which validates a number of assumptions used in the recent constitutive modeling attempts [6]. Comparing to the polymer constitutive models, the relation $N_1 = 2\tau^2/G$, where $G$ is the shear modulus, is widely used in polymer rheology. The above results show that it holds for yield stress fluids also, with $\psi_1 = 2/G$. The G obtained from this relation is of the same order as G' from the small-amplitude oscillatory measurements, as shown in Table 1; the two gels and the emulsion differ by about the same percentage, but in opposite directions merits further scrutiny. However in general $N_1 = 2\tau^2/G'$ is a very useful approximation if no normal stress data are available on a given system.

The final question is what the effect of the normal stresses is on the yielding. The criterion for yielding in stress space is given by the von Mises criterion; for a simple shear flow, with flow in the 1-direction, flow gradient in the 2-direction, and vorticity in the 3-direction, this criterion becomes:

$$N_1^2 + N_2^2 + (N_1 + N_2)^2 + 6\tau^2 \geq 2\sigma_y^2 \,.$$

Here, $\sigma_y$ is the elongational yield stress that would be measured in a shear-free uniform uniaxial deformation. This von Mises criterion has been tested for some yield stress materials [26] [27], but these studies do not account for normal stresses that are undoubtedly present. However our results here show that the normal stress terms, which scale as $\tau^4$, are much smaller than the shear stress term near yielding because of the small prefactors $\psi_{1,2}$. Our results then show that the normal stresses have a negligible effect (always less than 1%) on the yield criterion. This explains why the earlier experiments found good agreement with the Von Mises criterion. It also directly shows that for elongational flows, $\sigma_y \cong \sqrt{3} \ \tau_y$ and so our results on the normal stresses allow us to evaluate the elongational yield stress as well.

In conclusion, we have presented the first detailed study of the first and second normal stress differences for typical yield stress fluids, both above and below the yielding transition. We show that for these simple (non-thixotropic) yield stress fluids a coherent picture emerges, with a positive $N_1$ and a negative $N_2$, both of which vary quadratically with the shear stress in both the unyielded and the yielded states. Furthermore, the normal stresses do not go to zero when the shear rate does; a normal yield stress exists, as logically follows from the relation between shear and normal stresses found here. Besides the importance of finally being able to accurately measure these quantities, our results enable evaluation of the simple visco-elastic models that are at the basis of recent developments in the derivation of constitutive relations, which opens the way for their further refinement. Finally, our results allow one to understand when a yield stress fluid will flow in an arbitrary flow field, a problem of considerable practical importance since yield stress materials are typically processed in non-viscometric flows.

# The normal yield stress : supplementary information

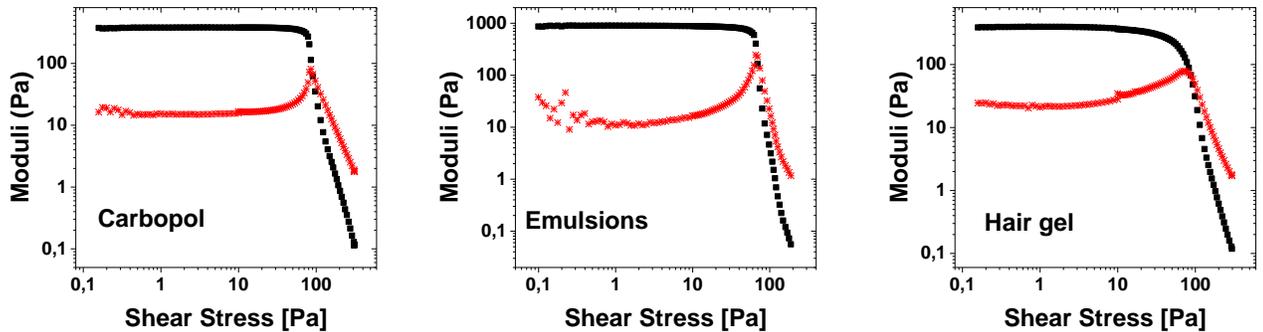

Figure A. Storage and loss moduli for cone- plate oscillation measurements on the three yield stress fluids. These measurements were done at a frequency of 0.1 Hz varying the amplitude of the stress oscillation.

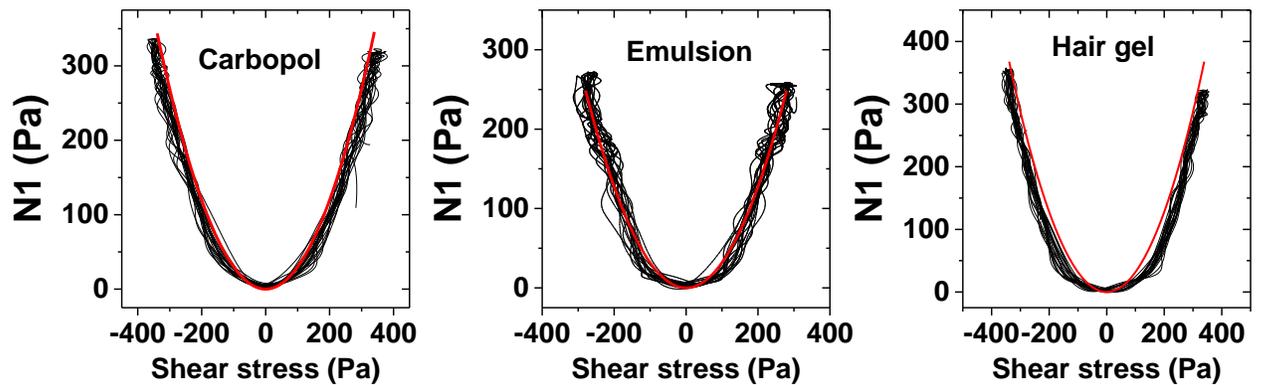

Figure B: Lissajous curves of normal stress versus shear stress for the different yield stress materials at high shear stress, with a cone plate geometry (CP 50) . The black curve represents the experimental data, the red curve the quadratic fit obtained from figure 2.

| Sample | $\tau_y$ (Pa) | K (Pa.s$^\beta$) | $\beta$ | G' plateau (Pa) |
|---|---|---|---|---|
| Emulsion | 53.56 | 27 | 0.42 | 900 |
| Hair gel | 56.5 | 18 | 0.44 | 400 |
| Carbopol | 65.51 | 28.13 | 0.40 | 380 |

Table A: Values of the Herschel-Bulkley parameters obtained from the fits to the flow curves plotted on figure 1 and value of the storage modulus G' before flowing (figure A SI)

## Regarding the contribution of the non-linear part of the signal to the shear rate.

The shear rate was initially derived from the shear stress by simply multiplying the amplitude of the shear stress by the oscillating frequency. This method is exact in the case of a perfectly sinusoidal signal. However, as the applied shear stress increases, the signal deviates from linearity. We therefore analyzed the shape of the signal to see what would be the influence of the third harmonic on the value of the shear rate. On figure C we plotted the FFT results of the strain signal for a hair gel sheared at 280Pa. The amplitude of the third harmonic is equal to 3% of the amplitude of the first harmonic. This leads to a contribution of 9% of the third harmonic to the shear rate. We therefore corrected the value of the shear rate with the contribution of the third harmonic every time it involved a change of 5% or more.

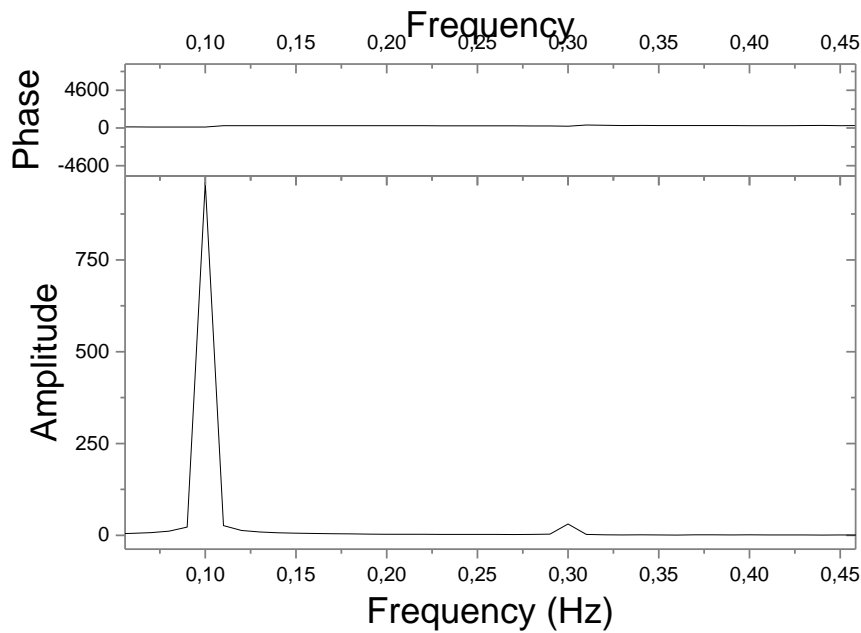

Figure C : Fast Fourier transform of the shear strain signal for a hair gel sheared at 280Pa

i